# A Performance Analysis Model of TCP over Multiple Heterogeneous Paths for 5G Mobile Services

**Jiayang Song[1], Ping Dong[1], Huachun Zhou[1], Tao Zheng[1], Xiaojiang Du[2], Mohsen Guizani[3]**

[1]  School of Electronic Information and Engineering, Beijing Jiaotong University, P. R. China;
    12111011@bjtu.edu.cn (J.S.); hczhou@bjtu.edu.cn (H.Z.); zhengtao@bjtu.edu.cn (T.Z.)
[2]  Dept. of Computer and Information Sciences, Temple University, USA; dxj@ieee.org
[3]  Dept. of Electrical and Computer Engineering, University of Idaho, Moscow, Idaho, USA;
    mguizani@gmail.com

**Abstract:** Driven by the primary requirement of emerging 5G mobile services, the demand for concurrent multipath transfer (CMT) is still prominent. Yet, multipath transport protocols are not widely adopted and TCP-based CMT schemes will still be in dominant position in 5G. However, the performance of TCP flow transferred over multiple heterogeneous paths is prone to the link quality asymmetry, the extent of which was revealed to be significant by our field investigation. In this paper, we present a performance analysis model for TCP over multiple heterogeneous paths in 5G scenarios, where both bandwidth and delay asymmetry are taken into consideration. The evaluation adopting parameters from field investigation shows that the proposed model can achieve high accuracy in practical environments. Some interesting inferences can be drawn from the proposed model, such as the dominant factor that affect the performance of TCP over heterogeneous networks, and the criteria of determining the appropriate number of links to be used under different circumstances of path heterogeneity. Thus, the proposed model can provide a guidance to the design of TCP-based CMT solutions for 5G mobile services.

## 1. Introduction

For emerging and promising 5G mobile services, despite their diverse application scenarios, it is widely agreed that they share a common primary requirement: either high data rate or high reliability. To meet such requirement, evolving wireless techniques and novel network infrastructures for 5G are no doubt necessary. However, we believe that the existing Concurrent Multipath Transfer (CMT) technology could also contribute to the fulfillment of needs of 5G mobile services since it can not only improve communication throughput, but also provide communication reliability. CMT in 5G scenarios will pool multiple heterogeneous wireless resources by employing a variety of Ratio Access Technologies (RATs) concurrently. Thus, the bandwidth of every RAT will be aggregated, achieving higher throughput. Also, thanks to diversity gain of heterogeneous RATs, the communication reliability can be improved. Meanwhile, it is potentially more viable to adopt CMT for mobile services in 5G since 5G is envisioned to consist of various types of RATs (such as millimeter wave communication, LTE-A and Wi-Fi), while more and more mobile devices have been equipped with multiple wireless interfaces [1].

Multipath techniques that can achieve CMT are still in development, while TCP-based CMT solutions will be in the dominant position. There are many reasons why multipath is not widely used. First, they cannot be widely applied to a variety of network environments. For example, the performance of MPTCP [2], the most popular multipath protocol working at a transport layer, will be severely degraded in some cases [3,4]. Second, the vast majority of operating systems, such as Windows, Linux, and MacOS, do not support multipath protocols well. Since most mobile services will still use TCP for now and for the foreseeable future, feasible CMT solutions for 5G services will be based on TCP. These solutions can be viewed as a middleware between the transport layer and network layer, which is transparent to the existing operating systems. Also, the interoperability between existing TCP based network infrastructure will not be compromised.

However, the performance of TCP flow transferred over multiple heterogeneous wireless networks would be adversely affected by path heterogeneity. This will be a critical feature of the highly integrative 5G system. Briefly, such performance degradation is due to the packet reordering issue [5] caused by the different link quality of employed heterogeneous wireless networks. The inherent re-sequencing mechanism of TCP can correct the problem when the packet reordering is no more than two positions [6]. However, the throughput may drop drastically due to the reduction of the TCP transmission window caused by more serious packet reordering [1]. Some contributions were proposed to solve the problem. Earliest Delivery Path First (EDPF) [7] schedules packets over different links based on their estimated delivery time. DAPS [8] distributes packets over different links depending on the ratio $RTT_{slow}/RTT_{fast}$ and $cwnd$. Yet, without the thorough understanding of TCP performance in the given situation, these contributions only provide limited improvement.

If we can analyze how heterogeneous networks affect the performance of TCP flow concurrently transferred over them, more efficient and elegant CMT schemes for 5G mobile services can be developed based on TCP and TCP-like congestion control protocols. Such TCP-based CMT schemes would be more deployable in 5G heterogeneous wireless networks since they are compatible with the current Internet infrastructure.

In this paper, a performance analysis model for TCP over multiple heterogeneous wireless networks is presented. To the best of our knowledge, no similar model has been reported in the literature. The proposed model can provide guidance to the design of novel CMT solutions for 5G mobile services. The main contributions of this paper are as follows:

(1) We have taken field investigation on present heterogeneous wireless networks to reveal the severe extent of link quality asymmetry in terms of delay and bandwidth. This proves that the impact of network heterogeneity in future 5G is anything but empty talk.

(2) A performance analysis model is derived based on the careful analysis of segments transmission and acknowledgement response over multiple heterogeneous paths. Both bandwidth asymmetry and delay asymmetry are taken into consideration in the proposed model.

(3) High analytical accuracy is achieved by comparison to the simulation using parameters from field investigation. It proves that our model can be applied in practical environments. Simulation of TCP over multiple heterogeneous paths is created in NS3, and the predicted throughput using the proposed model can fit the simulation results with high accuracy.

(4) Some interesting inferences are drawn from the proposed model. First, compared to bandwidth asymmetry, delay asymmetry between multiple links is the dominant factor that affects the performance of TCP over heterogeneous paths. Second, the criteria of determining the appropriate number of links to be employed to optimize the TCP multipath performance is discussed.

The remainder of this paper is organized as follows. Some related work is introduced in Section 2. Section 3 details the issue of link quality asymmetry based on the results of field investigation. In Section 4, the performance analysis model for TCP over heterogeneous paths are derived. The accuracy of the proposed mode is shown in Section 5. In Section 6 we investigate the effect of path heterogeneity based on the proposed model. Section 7 concludes the paper.

## 2. Related Work

To meet the requirement for high data rate and reliability, some contributions were proposed to try to achieve stable and high-quality communication based on multipath transmission. SCTP [9,10] and its extensions [11,12] try to aggregate the bandwidth of multiple paths. MPTCP [13], a multipath extension to TCP, has also been standardized to transmit data over multiple paths simultaneously to improve reliability and throughput. IETF Multiple Interfaces (MIF) working group is developing the standards [14] for nodes with multiple interfaces. Besides these papers, there are some other works (e.g., [15-18]) studied security related networking issues, especially the key management topics [19,20].

Recently, the cellular-based solutions are generating more interest with the rapid development of 5G heterogeneous networks. For example, femocells-based schemes [21,22] were proposed to support seamless mobility and maximize the network recourse utilization using multiple interfaces.

However, apart from the practical deployment challenges, such as the existence of various types of middle boxes [3], the main difficulty is that the performance of multipath solutions may decrease significantly under the circumstances of path heterogeneity, especially when there are some bottleneck paths [4,23-25].

Packet reordering is considered the dominant challenge for multipath transmission because it leads to an undesirable reduction in throughput [1]. RFC5236 [26] introduces a metric named reorder density to show how far packets are displaced from their original position. Therefore, an efficient multipath solution must reduce the impact of packet reordering to alleviate its effects.

Multipath forwarding is the main reason of packet out-of-order [27]. Different technologies and different paths can lead to significant differences in delay and bandwidth. When packets are forwarded over paths with different characteristics, they are likely to arrive at the receiver out of order.

Some state-of-art [28,29] has measured the characteristics of heterogeneous paths in terms of delays. However, their main purpose is to analyze the performance of different scheduling algorithms in heterogeneous networks, rather than theoretically analyze the relationship between path diversity and TCP performance.

The research of TCP performance analysis, especially in terms of throughput, is still making progress, as TCP is one of most widely deployed transport protocols in today's Internet. The research can be categorized into two kinds: one aims at improving the accuracy of prior model by novel methods [30-32], the other focuses on the performance of TCP applied in emerging scenarios [33,34]. However, the proposed models in these papers only analyze the situation where single path is used for transmitting TCP segments.

Overall, to the best of our knowledge, no one has given a performance analysis model to analyze TCP performance over multiple paths with different link quality in heterogeneous networks, although there are many schemes [35] working at different protocol layers that are proposed to try to improve the performance over multiple paths. We believe that this model can help us design more practical multipath schemes in the future wireless networks.

## 3. Problem Description and Network Model

Network heterogeneity will become a concrete issue in 5G with the popularity of multi-access devices and deployment of emerging heterogeneous RATs. Multi-access devices that can connect to more than one wireless networks are gaining bigger market share, such as smart phones supporting dual-SIM dual stand-by mode. These devices can concurrently use up to three interfaces, including Wi-Fi, for data transmission. For such a device, the connected multiple wireless networks may share heterogeneous access technologies (e.g., WLAN vs. cellular network), heterogeneous standards (e.g., FDD-LTE vs. TD-LTE) or heterogeneous service providers. Even if two interfaces are connected to an identical wireless network, the wireless signals are likely to experience heterogeneous pass loss due to small scale fading. Considering that in 5G more heterogeneous RATs will be deployed and utilized by multi-access devices, the network heterogeneity issue will become more severe than in previous four generations.

Network heterogeneity of multi-access devices is intuitively revealed by the difference in network link quality. For two heterogeneous wireless networks, their network link quality is normally different from each other, to which we refer as network link quality asymmetry. Generally, Data Rate (DR) and Round-Trip Time (RTT) are used to describe the network link quality, for DR reveals the capacity of a network link, while RTT directly reflects the transmission delay. Accordingly, the network link quality asymmetry can be indicated by DR asymmetry and RTT asymmetry.

Intuitively, the performance of TCP transmission would be prone to network link quality asymmetry, if multiple heterogeneous wireless networks are concurrently employed for delivering segments, they will consequently degrade the performance of TCP-based CMT in 5G. This is because the transmitted segments would suffer different transmission delays due to the dissimilar network link quality of employed wireless networks. This results in segments reaching the receiver out-of-order. This segment reordering issue is widely regarded as the major challenge that undermines the performance of concurrent multipath transmission, as it causes unnecessary retransmission, prevents the congestion window from growing and disrupt ACK-clocking. The higher network link quality asymmetry becomes, the more negative impact it has on TCP performance. The analytical discussion of relationship between the performance of TCP over multiple wireless networks and the link quality asymmetry will be detailed in section IV.

To investigate the extent of network link quality asymmetry in real-world situation, we have taken a filed measurement on a group of heterogeneous wireless networks and found that their link quality deviated significantly from each other. The measurement was carried out in a test train running on a newly constructed high-speed railway before its service, where few passengers were on board, to eliminate the interference from other wireless devices. Inside the test train, a dedicated box PC with our proprietary measuring program was deployed to automatically measure and store the download DR and RTT of a certain wireless network. Incorporating different kinds of wireless modems, this device can simultaneously access multiple heterogeneous mobile networks. In the measurement, up to eight modems were adopted, including three FDD-LTE modems of China Telecom (CT), three FDD-LTE modems of China Unicom (CU) and two TD-LTE modems of China Mobile (CM). After the measurement, a group of RTT dataset and two download DR values (average and maximum) were collected on each modem.

The statistics from the measurement result is shown in Figure 1. Regarding RTT, a boxplot diagram is depicted based on collected dataset of each modem. The rectangle in a boxplot diagram represents the interquartile ranges (IQR) of the variation, while the segment inside the rectangle represents the median. By visually comparing the two boxplot diagrams, statistical inference can be made about the difference of two dataset. If the median of one dataset does not overlap the IQR of the other dataset, it can be inferred that difference exists between two datasets. Further, if two IQRs don't overlap, the difference is significant. Applying this criterion to Figure 1, we can infer that the RTT of CT1, CU3, CM1 and CM2 are significantly higher than those of CT2, CU1, CU2. Meanwhile, the RTT of CT1, CU3 and CM1 are different from the others. These conclusions can reveal the dispersion of RTT among eight modems.

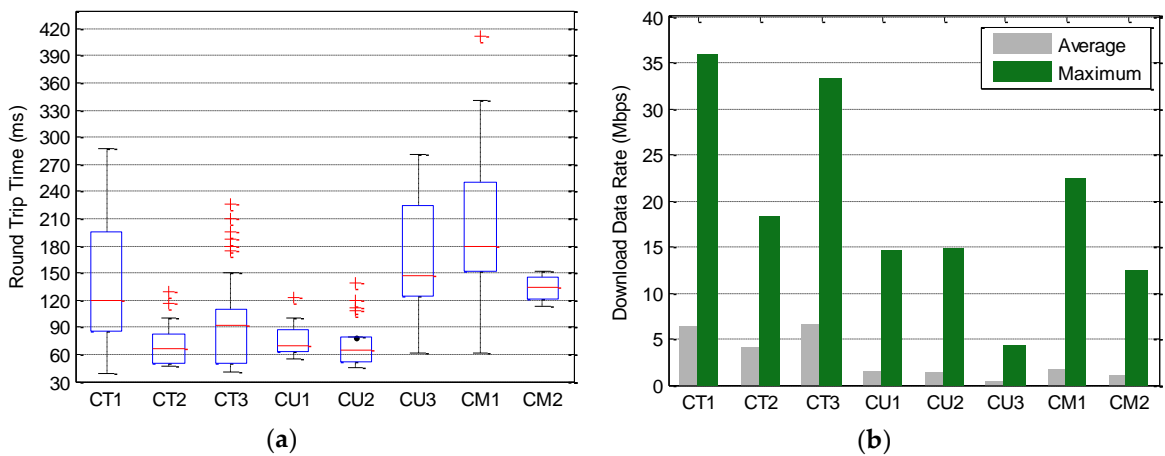

**Figure 1.** Results of field measurement regarding link quality asymmetry of heterogeneous wireless networks. CT1, CT2 and CT3 are FDD-LTE of China Telecom, CU1, CU2 and CU3 are FDD-LTE of China Unicom, CM1 and CM2 are TD-LTE of China Mobile. (a) depicts the boxplot RTT statistics, (b) shows the maximum and average download data rate, both of which can reveal the significant difference in link quality of heterogeneous wireless networks.

Regarding download DR, the average and maximum values are shown using bar graphs. For maximum download DR, the ratio between the highest (CT1) and the lowest (CU3) is 8.2. As for average download DR, this ratio is even more pronounced, reaching 15.1. This means that notable deviation exists in download DR among different modems.

To sum up, the field measurement results allow to conclude that the network link quality asymmetry in real-world situation is truly significant. Besides, it is revealed that the link quality asymmetry not only exists between two heterogeneous networks, but also between two modems using access technology operated by same telecommunication company. According to above conclusions, we can infer that the network heterogeneity in future 5G will be more severe and become a concrete threat, since the wireless networks in 5G will become more diverse than nowadays with the deployment of emerging RATs.

As we have demonstrated, the network heterogeneity will affect the performance of TCP-based CMT solutions for 5G mobile services. Thus, it is very essential to create a quantitative performance analysis model regarding the relationship between the link quality asymmetry and TCP multipath performance. To build such a performance analysis model, we first present the network model of TCP flow transferred over multiple heterogeneous links, as shown in Figure 2. In this network model, the segments of single TCP connection are concurrently distributed over multiple paths between two endpoints. We use $L = \{l_1, l_2, \ldots, l_n\}$ to denote the set of $n$ available heterogeneous links, $D = \{d_1, d_2, \ldots, d_n\}$ to denote the set of round-trip propagation delay, and $B = \{b_1, b_2, \ldots, b_n\}$ to denote the set of bandwidth. The bandwidth and round-trip propagation delay of link $l$ is $b_l$ and $d_l$ To simplify the analysis, we assume that the propagation delay from the receiver to sender is zero. Round Robin (RR) is used to dispatch packets in the given network mode, which let multiple paths take turns in transferring data packets in a periodically repeated order. We choose NewReno [36] as the congestion control algorithm since it is still the widely deployed version of TCP.

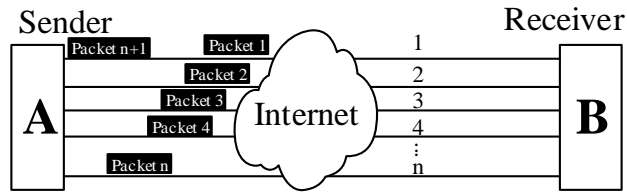

**Figure 2.** The network model of TCP over multiple heterogeneous paths

## 4. Performance Analysis Model

In this section, the performance analysis model of TCP over multiple heterogeneous paths is built by analyzing the average throughput. We divide the TCP flow into consecutive transmission round. The duration time as well as the number of segments transmitted at each round are first analyzed. Then, the average throughput is derived using an iteration model. At last, the effect of link quality asymmetry on average throughput is discussed. Table 1 summarizes important parameters used in this paper.

### 4.1. Analysis of i-th Transmission Round

First, we focus on the transmission of segments at sender side. Let i denote the number of transmission round from the beginning of the transmission. At i-th round, sender transmits a certain number of unsent segments and waits for the acknowledgements. Since in most TCP implementations (such as NS3) only non-duplicate ACK triggers the transmission of previously unsent data. We can conduct that the i-th round begins with the arrival of i-th non-duplicate ACK.

**Table 1.** Notations.

| Parameter | Description |
|---|---|
| $L$ | The set of available links |
| $n$ | The number of available links |
| $D$ | The set of round-trip propagation delay of available links |
| $B$ | The set of bandwidth of available link |
| $s$ | The size of a segment |
| $m_{ACK}$ | Receiver reply an ACK after receiving $m_{ACK}$ consecutive segments |
| $SGM_{i,j}$ | The j-th segment that sender transmits at i-th round |
| $w_i$ | The congestion window of i-th round |
| $\delta_w^i$ | The increment of congestion window at i-th round |
| $A_i$ | The number of segments acknowledged by i-th non-duplicate ACK |
| $C_i$ | The number of segments that can be transmitted at i-th round |
| $T_i$ | The time between the i-th round and (i+1)-th round |
| $\eta_{i,j}$ | The number of the link used to transmit the j-th segment of $C_i$ at i-th round |
| $D_{i,j}$ | The propagation delay and queuing delay of j-th segment of $C_i$ at i-th round |
| $I_s$ | The number of rounds that the slow start phase ends |
| $W_s$ | The slow start threshold of congestion window |
| $W_I$ | The initial value of congestion window |

Let $C_i$ denote the total number of segments transmitted at $i$-th round. $C_i$ equals the free space in the congestion window, which is composed of two parts: the increment in size of congestion window and the decrement in number of outstanding segments. We define $w_i$ as the size of the congestion window of $i$-th round, and $\delta_w^i = w_i - w_{i-1}$ as the increment of the congestion window. Let $A_i$ denote the number of segments newly acknowledged by $i$-th non-duplicate ACK, then $C_i$ can be expressed as:

$$C_i = (A_i + \delta_w^i).\tag{1}$$

The $j$-th segment of $C_i$ is defined as $SGM_{i,j}$. Let $\eta_{i,j}$ denote the number of the link used to send the $SGM_{i,j}$, where $l_{\eta_{i,j}} \in L$ and $\eta_{i,j} \in \{1, 2, \dots, n\}$. Supposing segments are scheduled over $n$ links in a round-robin manner, and the first one travels over link $l_1$. Hence $\eta_{i,j}$ can expressed as:

$$\eta_{i,j} = \left[ \left( j - 1 + \sum_{k=1}^{i-1} C_k \right) \bmod n \right] + 1.\tag{2}$$

The round-trip propagation delay as well as the bandwidth of link $l_{\eta_{i,j}}$ are $d_{\eta_{i,j}}$ and $b_{\eta_{i,j}}$ respectively. Let $D_{i,j}$ be the time elapsed between the beginning of $i$-th round and when $SGM_{i,j}$ reaches the receiver, which is the sum of queuing delay and propagation delay experienced by $SGM_{i,j}$. Thus,

$$D_{i,j} = \frac{\left( \left\lfloor \frac{j}{n} \right\rfloor + 1 \right) s}{b_{\eta_{i,j}}} + d_{\eta_{i,j}}.\tag{3}$$

In (3), $\lfloor j/n \rfloor$ is the quotient of $j$ and $n$, while $S$ is the average size of segments. The queuing delay is represented by $\left[ \left( \left\lfloor \frac{j}{n} \right\rfloor + 1 \right) s \right] / b_{\eta_{i,j}}$, while $d_{\eta_{i,j}}$ represents the propagation delay.

Then we discuss the arrival of segments and the response of ACKs at receiver side. We define $T_i$ as the latency between the beginning of $i$-th round and the time when sender receives the first non-duplicate ACK that starts the $(i + 1)$-th round from receiver. The number of segments the first non-duplicate acknowledges is exactly $A_{i+1}$. A non-duplicate ACK will be fired by receiver only if: 1) an expected number of consecutive segments are received, 2) the first out-of-order segments arrives after some consecutive segments or 3) a segment that fills the gap in the receiver's buffer arrives. The satisfaction of these criteria highly associates with the arrival order of the first segment transmitted at $i$-th round, which is $SGM_{i,1}$. Hence, based on whether $SGM_{i,1}$ is the first to reach the receiver, we respectively calculate $T_i$ and $A_{i+1}$.

### 4.1.1. Case I: $SGM_{i,1}$ is the first to reach the receiver

We define $P(F)$ as the probability that $SGM_{i,1}$ arrives at the receiver first, which can be presented as:

$$P(F) = P\left(D_{i,1} = \min_{j \in \{1,2,...,C_i\}}\{D_{i,j}\}\right). \tag{4}$$

The segments are scheduled over the links in a round-robin manner, thus $D_{i,j}$ follows a uniform distribution after a large amount of transmission rounds. Hence $P(F)$ approximately equals $1/C_i$.

Most TCP implementations (such as NS3) utilize a counter to delay replying cumulative ACK. Let $m_{ACK}$ denote this counter, after receiving $m_{ACK}$ consecutive segments receiver will reply an ACK. In this case, since the receiver receives $SGM_{i,1}$ first, it will wait for the following $m_{ACK} - 1$ segments before replying an ACK until the arrival of first out-of-order segment, as shown in Figure 3. Let $m$ be the number of consecutive segments received before the arrival of first out-of-order segments. In other words, $SGM_{i,2}$ to $SGM_{i,m}$ arrive consecutive and $SGM_{i,m+1}$ is out of order. Thus, receiver will reply the first non-duplicate ACK acknowledging $m$ segments approximately after the arrival of $SGM_{i,m}$.

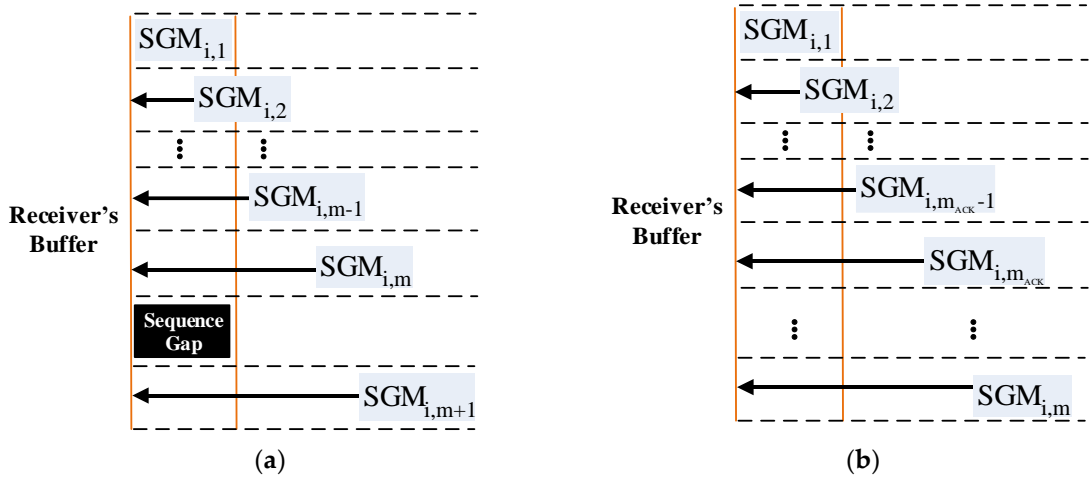

**Figure 3.** Case I: $SGM_{i,1}$ is first to reach the receiver, and the consecutive received segments may be smaller (a) or larger (b) than $m_{ACK}$

If $m$ is smaller than $m_{ACK}$, we have $T_i \cong D_{i,m}$ and $A_{i+1} = m$, where $D_{i,m}$ is the time between the beginning of $i$-th round and arrival of $SGM_{i,m}$. The probability $P(m < m_{ACK} \mid F)$ can be calculated as:

$$P(m < m_{ACK} \mid F) = \sum_{k=1}^{m_{ACK}-1} \frac{(C_i - k - 1)(C_i - k - 1)!}{(C_i - 1)!}. \tag{5}$$

If m is equal to or larger than $m_{ACK}$, $T_i \cong D_{i,m_{ACK}}$ and $A_{i+1} = m_{ACK}$. The corresponding probability $P(m \geq m_{ACK} \mid F)$ can be calculated as:

$$P(m \geq m_{ACK} \mid F) = \frac{1}{(C_i - 1)!} + \sum_{k=m_{ACK}}^{C_i-2} \frac{(C_i - k - 1)(C_i - k - 1)!}{(C_i - 1)!}. \tag{6}$$

Let $E'(T_i)$ and $E'(A_{i+1})$ denote the expected value of $T_i$ and $A_{i+1}$ under the condition that $SGM_{i,1}$ is the first to reach the receiver. Based on the probabilities calculated in (5) and (6), and the corresponding $T_i$ and $A_{i+1}$, $E'(T_i)$ and $E'(A_{i+1})$ can be derived as:

$$E'(T_i) = \sum_{k=1}^{m_{ACK}-1} D_{i,k} \frac{(C_i - k - 1)(C_i - k - 1)!}{(C_i - 1)!}$$
$$+ D_{i,m_{ACK}} \sum_{k=m_{ACK}}^{C_i-2} \frac{(C_i - k - 1)(C_i - k - 1)!}{(C_i - 1)!} + \frac{D_{i,m_{ACK}}}{(C_i - 1)!'} \tag{7}$$

$$E'(A_{i+1}) = \sum_{k=1}^{m_{ACK}-1} k \frac{(C_i - k - 1)(C_i - k - 1)!}{(C_i - 1)!}$$
$$+ m_{ACK} \sum_{k=m_{ACK}}^{C_i-2} \frac{(C_i - k - 1)(C_i - k - 1)!}{(C_i - 1)!} + \frac{m_{ACK}}{(C_i - 1)!}. \tag{8}$$

### 4.1.2. Case II: $SGM_{i,1}$ is not the first to reach the receiver

$P(\bar{F})$ is defined as the probability of the $SGM_{i,1}$, where it is not the first to reach the receiver, which approximately equals $(1 - 1/C_i)$. In this case, the receiver will not reply any non-duplicate ACK before the arrival of $SGM_{i,1}$. Moreover, since the segments transmitted later than when $SGM_{i,1}$ arrives at the receiver earlier than itself, there must be gaps in the receiver's buffer before the arrival of $SGM_{i,1}$. As shown in Figure 4, the receiver will immediately reply a non-duplicate ACK after receiving $SGM_{i,1}$, since the $SGM_{i,1}$ will fill part of the existing gap. Hence, $T_i$ equals $D_{i,1}$.

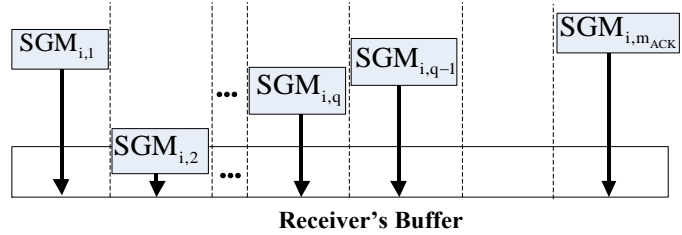

**Figure 4.** Case II: $SGM_{i,1}$ is not the first to reach the receiver

Let $(q - 1)$ denote the number of received consecutive segments counting from $SGM_{i,2}$ before the arrival of $SGM_{i,1}$. Hence, the number of segments the non-duplicate ACK can acknowledge equals $q$, consequently we have $A_{i+1} = q$. The probability of $P(q = k \mid \bar{F})$ is derived as follows:

$$P(q = k \mid \bar{F}) = \begin{cases} \dfrac{(C_i - 2)(C_i - 1)!}{2C_i! - 2(C_i - 1)!}, & k = 1 \\[2mm] \dfrac{(C_i - 1)!}{C_i! - (C_i - 1)!}, & k = C_i \\[2mm] \dfrac{(C_i - k - 1)\left[\sum_{l=k+1}^{C_i} \dfrac{(l - 2)!}{(l - k - 1)!}(C_i - l + 1)\right]}{C_i! - (C_i - 1)!}, & 1 < k < C_i \end{cases} \tag{9}$$

Let $E''(T_i)$ and $E''(A_{i+1})$ denote the expected value of $T_i$ and $A_{i+1}$ under the condition that $SGM_{i,1}$ is not the first to reach the receiver, which can be derived as:

$$E''(T_i) = D_{i,1}, \tag{10}$$

$$E''(A_{i+1}) = \frac{(3C_i - 2)(C_i - 1)!}{2C_i! - 2(C_i - 1)!} + \sum_{k=2}^{C_i-1} k \frac{(C_i - k - 1)\left[\sum_{l=k+1}^{C_i} \dfrac{(l - 2)!}{(l - k - 1)!}(C_i - l + 1)\right]}{C_i! - (C_i - 1)!}. \tag{11}$$

Based on case I and case II, we can derive the expected value of $T_i$ and $A_{i+1}$ as $E(T_i) = E'(T_i) P(F) + E''(T_i)P(\bar{F})$ and $E(A_{i+1}) = E'(A_{i+1}) P(F) + E''(A_{i+1})P(\bar{F})$. Given that $P(F) = 1/C_i$ and $P(\bar{F}) = (1 - 1/C_i)$, $E(T_i)$ and $E(A_{i+1})$ can be calculated as follows:

$$E(T_i) = E''(T_i) + \frac{1}{C_i}\big(E'(T_i) - E''(T_i)\big), \tag{12}$$

$$E(A_{i+1}) = E''(A_{i+1}) + \frac{1}{C_i}\big(E'(A_{i+1}) - E''(A_{i+1})\big). \tag{13}$$

From (8) and (11) we can find that the expected value of $A_{i+1}$ depends on $C_i$, which means $A_{i+1}$ is a function of $C_i$. For simplicity, we define $A_{i+1} = F(C_i)$. Since $C_{i+1}$ equals the sum of $A_{i+1}$ and $\delta_w^{i+1}$, the increment of the congestion window needs to be discussed.

In the slow start phase, the congestion window is incremented by one segment for each ACK, thus $\delta_w^i$ equals 1. Let $W_s$ denote the slow start threshold of congestion window, and $W_I$ the initial size of congestion window. Let $I_s$ be the number of rounds that the slow start phase ends. Since the congestion window is increased by one every round, thus:

$$I_s = W_s - W_I. \tag{14}$$

In the congestion avoidance phase, the congestion is increased by $1/w$ on every incoming ACK that acknowledges new data. Thus, we have $\delta_w^i = 1/w_{i-1}$. The congestion window at $(i + 1)$-th round can be expressed as:

$$w_{i+1} = \begin{cases} w_i + 1, i < I_s \\ w_i + \dfrac{1}{w_i}, i \geq I_s \end{cases} \tag{15}$$

Based on the above analysis, the relationship between $C_{i+1}$ and $C_i$ can be derived as (15), where function $F(\cdot)$ is defined in (13):

$$C_{i+1} = \begin{cases} F(C_i) + 1, i < I_s \\ F(C_i) + \dfrac{1}{w_i}, i \geq I_s \end{cases} \tag{16}$$

### 4.2. Iteration for Average Throughput

According to (16), the number of segments transmitted at next round can be derived based on that at current round. Thus, the total segments transmitted from the beginning to current round of transmission can be calculated by iteration from the first round. The total time spent on transmitting can also be obtained by summing up the duration time of each transmission round. Consequently, the average throughput can be derived.

Therefore, we formulate the performing process of the model iteration as follows:

*Step* **1**: Supposing $E$ bytes of data are expected to be received by the receiver. At the first round of transmission, $C_1$ segments are sent within $E(T_1)$ seconds, where $C_1$ equals the initial size of the congestion window, which is $W_I$. $E(T_1)$ can be calculated according to (12).

*Step* **2**: At $i$-th round ($i \geq 2$), substituting $w_{i-1}$ into (15), we can get $w_i$.

*Step* **3**: At $i$-th round ($i \geq 2$), substituting $C_{i-1}$ and $w_{i-1}$ into (16), we can get $C_i$. Further, according to (12), $E(T_i)$ is computed.

*Step* **4**: Compute total transmitted bytes from beginning to $i$-th round, which is:

$$Total\ Transmitted\ Bytes = s\sum_{k=1}^{i} C_k. \tag{17}$$

*Step* **5**: Let $\hat{T}$ denote total transmission time from beginning to $i$-th round, which can be computed as:

$$\hat{T} = \sum_{k=1}^{i} E(T_k). \tag{18}$$

**Step 6**: If total transmitted bytes is smaller than $E$, which is the number of bytes expected by the receiver, repeat *Step* 2-5. Otherwise, the iteration stops, and the average throughput can be calculated as:

$$Average\ Throughput = \frac{E}{\hat{T}}. \tag{19}$$

*4.3. Discussion of Link Quality Asymmetry*

Here we discuss how link quality asymmetry affects the throughput of TCP transferred over multiple heterogeneous links. According to the proposed model, when transmitting a certain number of bytes, the average throughput is inversely proportional to the total transmission time $\hat{T}$. Since $\hat{T}$ is the sum of $E(T_i)$ defined in (12), the average throughput decreases with increasing $E(T_i)$. Substituting $E'(T_i)$ defined in (7) and $E''(T_i)$ defined in (10) into (12), $E(T_i)$ can be evaluated as:

$$E(T_i) = \sum_{k=m_{ACK}}^{c_i-2} (D_{i,m_{ACK}} - D_{i,1}) \frac{(C_i-k-1)(C_i-k-1)!}{C_i!} + \frac{(D_{i,m_{ACK}} - D_{i,1})}{C_i!} + D_{i,1}$$
$$+ \sum_{k=1}^{m_{ACK}-1} (D_{i,k} - D_{i,1}) \frac{(C_i-k-1)(C_i-k-1)!}{C_i!}. \tag{20}$$

Apart from parameter $D_{i,1}$, $E(T_i)$ is primarily associated with $(D_{i,k} - D_{i,1})$, where $k \in [1, m_{ACK}]$. $(D_{i,k} - D_{i,1})$ represents the time difference between the arrival of the first transmitted segment $SGM_{i,1}$ and the $k$-th transmitted segment $SGM_{i,k}$ at the receiver side. Longer time difference increments the overall $E(T_i)$. Substituting (3), $(D_{i,k} - D_{i,1})$ can be evaluated as:

$$D_{i,k} - D_{i,1} = \left(d_{\eta_{i,k}} - d_{\eta_{i,1}}\right) + \frac{\left(b_{\eta_{i,1}} - b_{\eta_{i,k}}\right)s}{b_{\eta_{i,k}} b_{\eta_{i,1}}} + \frac{\left\lfloor \frac{k}{n} \right\rfloor s}{b_{\eta_{i,k}}}. \tag{21}$$

As demonstrated in (18), $(D_{i,k} - D_{i,1})$ is mainly dominated by two elements, $(d_{\eta_{i,k}} - d_{\eta_{i,1}})$ and $(b_{\eta_{i,1}} - b_{\eta_{i,k}})$. $(d_{\eta_{i,k}} - d_{\eta_{i,1}})$ and $(b_{\eta_{i,1}} - b_{\eta_{i,k}})$ are respectively the delay difference and bandwidth difference between two links that transmit $SGM_{i,1}$ and $SGM_{i,k}$, i.e., $l_{\eta_{i,1}}$ and $l_{\eta_{i,k}}$. Increasing $(d_{\eta_{i,k}} - d_{\eta_{i,1}})$ and $(b_{\eta_{i,1}} - b_{\eta_{i,k}})$ leads to greater $(D_{i,k} - D_{i,1})$, and consequently causes larger $E(T_i)$, which eventually results in a decrease in average throughput.

As mentioned earlier, when scheduled in round-robin manner, the possibility of selecting one of $n$ available links to transmit a certain segment follows a uniform distribution after a large amount of transmission rounds. Thus, $l_{\eta_{i,1}}$ and $l_{\eta_{i,k}}$ can represent any two links of set $\{l_1, l_2, ..., l_n\}$. Note that $l_{\eta_{i,1}}$ is not the first link of $n$ available links, but the link used to transmit $SGM_{i,1}$. Equally, $(d_{\eta_{i,k}} - d_{\eta_{i,1}})$ and $(b_{\eta_{i,1}} - b_{\eta_{i,k}})$ can be the delay difference and bandwidth difference between any two links. From this point of view, $(d_{\eta_{i,k}} - d_{\eta_{i,1}})$ and $(b_{\eta_{i,1}} - b_{\eta_{i,k}})$ reflect the extent of deviation in link quality of all links. We refer to such delay difference and bandwidth difference between any two links as delay asymmetry and bandwidth asymmetry. Therefore, it can be concluded that the average throughput is subject to delay asymmetry and bandwidth asymmetry. The more significant these two parameters become, the lower average throughput will be.

To quantify delay asymmetry, we introduce Average Delay Asymmetry, which is defined as the average absolute delay difference between any two links of $n$ available links. Average Delay Asymmetry can be calculated as:

$$Average\ Delay\ Asymmetry = \frac{2\sum_{q=2}^{n}\sum_{r=1}^{q-1}|d_r - d_q|}{n(n-1)}, d_q, d_r \in D. \tag{22}$$

Similarly, Average Bandwidth Asymmetry can be defined as:

$$Average\ Bandwidth\ Asymmetry = \frac{2\sum_{q=2}^{n}\sum_{r=1}^{q-1}|b_r - b_q|}{n(n-1)}, b_q, b_r \in B. \tag{23}$$

Average Delay Asymmetry and Average Bandwidth Asymmetry can both affect the performance of TCP transferred over multiple heterogeneous links. Comparison of extent of these two parameters on TCP performance will be presented in section VI.

## 5. Simulation Study

The proposed model in section IV is evaluated by comparing its prediction with the results of simulation. To verify that our model can be used in practical environments, the parameters in both model prediction and simulation are taken from the datasets collected in the field measurement discussed in section III. The simulation of TCP over multiple heterogeneous links is implemented in Network Simulator 3 (NS3) [37].

### 5.1. Simulation Implementation

Figure 5 depicts the simulation topology. Two endpoints are connected by multiple Point-to-Point Protocol (PPP) links. At each point, apart from two PPP network adapters, a virtual network device (VND) that works at the network layer was added. An IP address is assigned to VND. Between two endpoints, a TCP connection binding to the IP addresses of two VNDs is established. At both endpoints, TCP NewReno is used. When a TCP segment of the established connection is pushed down to the network layer, the corresponding IP packet will be forwarded to VND. VND then passes the IP packet to a dedicated packet-processing program attached to it. The IP packet will be encapsulated into a UDP datagram and then sent to the peer from one of the PPP network adapters. A Round-Robin scheduling algorithm is employed to decide which network adapter will be used to transmit the subsequent encapsulated packets. Thus, the segments of the single TCP connection established between two endpoints will be concurrently transmitted from all the available PPP network adapters.

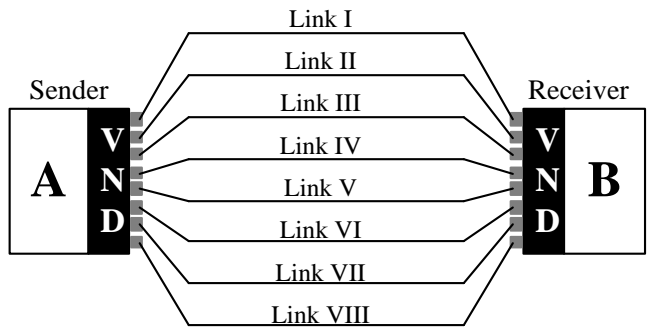

**Figure 5.** The simulation topology.

To measure the throughput of simulated TCP over multiple heterogeneous links in NS3, a sending application is installed on endpoint A, and a receiving application is installed on endpoint B. A then sends $C$ bytes data to B, and B records the transfer finish time as $T$ seconds. Thus, the throughput can be calculated as $C/T$ bytes per second.

### 5.2. Evaluation Methodology

Using the proposed model and the simulation respectively, two sets of throughputs of TCP over multiple heterogeneous links are obtained for comparison. For each case, the derivation of throughput is performed under different number of heterogeneous links employed for concurrent transmission. When utilizing a certain number of links, the bandwidth of a link remains constant but different from those of the other links. The value of delay of a link is fetched from an individual dataset associated with that link. For example, if $m$ links are employed for a concurrent transmission, and the delay dataset of each link contains $n$ values, then there will be $n^m$ combinations of delay values. The Average Delay Asymmetry of $n^m$ groups of delay will be calculated and sorted, from which 36 groups of delay will be evenly selected. For selected groups of delay values, the derivation of throughput is repeated using simulation and proposed model correspondingly.

### 5.3. Parameter Settings

The parameters for model prediction or simulation are taken from the measurement results of field investigation towards the wireless network heterogeneity, as described in section III. Since eight modems were measured during the investigation, up to eight links can be employed for concurrent transmission in model prediction or simulation, namely link I to link VIII. For example, if our links are needed, Link I, II, III and IV will be utilized. Link I, II and III represents FDD-LTE of China Telecom, Link IV, V and VI represents FDD-LTE of China Unicom, link VII and VIII represents TD-LTE of China Mobile. The bandwidth of link I to link VII are set as the maximum measured download data rates shown in Figure 1(b), which are respectively 35.9Mbps, 18.4Mbps, 33.3Mbps, 14.7Mbps, 14.8Mbps, 4.4Mbps, 22.5Mbps and 12.5Mbps.

The field measurement results of RTT of a modem are directly adopted as the delay dataset of corresponding link in the simulation.

The other parameters used in proposed model are set according to Table 2.

**Table 2.** Evaluation Parameters.

| Parameter | Value | Parameter | Value |
|-----------|-------|-----------|-------|
| $W_I$ | 536 bytes | $W_S$ | 65535 bytes |
| $S$ | 536 bytes | $m_{ACK}$ | 2 segments |

### 5.4. Evaluation Results

We introduce prediction accuracy to evaluate the proposed model's consistency to simulation results. Supposing the predicted throughput using the proposed model is $T_M$, the derived throughput using simulation under same circumstance is $T_S$, then prediction accuracy is defined as:

$$Prediction\ Accuracy = 1 - \frac{|T_S - T_M|}{T_S}. \tag{24}$$

The evaluation results with number of links varying from 2 to 8 are depicted in Figure 6, where throughput is plotted against the cyan circles, which represent the simulation results, and the red crosses indicate the predicted values using the proposed model. It can be observed that there is a good match between the model prediction and the simulation results in all cases. With the number of links employed for concurrent transmission varying from 2 to 8, the prediction accuracies are 89.68%, 83.14%, 79.26%, 75.99%, 73.24%, 71.06% and 69.50%. The prediction accuracies slightly drop as the number of utilized links increases. This is due to that the error introduced by the randomness becomes larger in the proposed mode when the number of links available for transmission grows. Even so, the average prediction accuracy can reach 77.41%.

Since the parameters of link quality (i.e., bandwidth and delay) used in the simulation are adopted from the results of field measurement, we can conclude that the proposed model is also accurate for TCP over multiple heterogeneous links in practical environment.

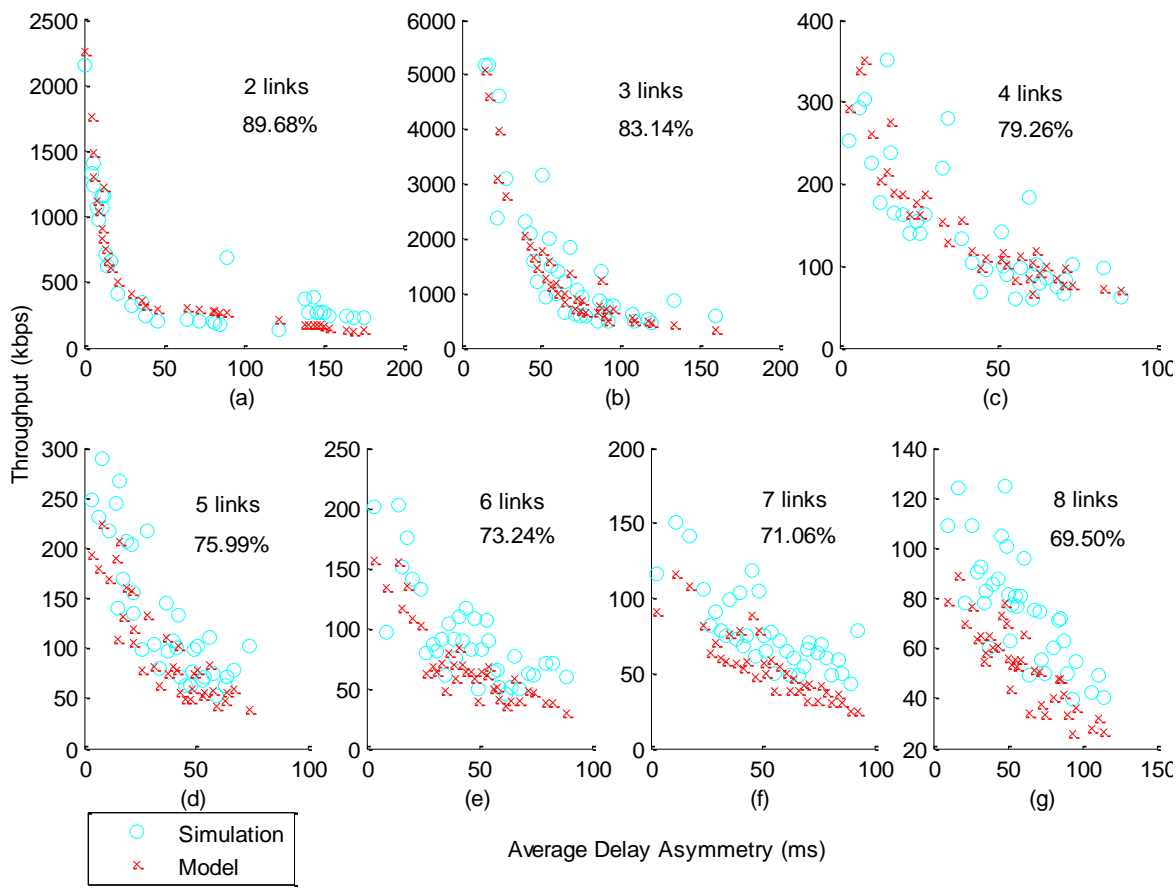

**Figure 6.** Comparison between the proposed model and simulation experiment. The number of links employed for concurrent transmission varies from 2 to 8, and the corresponding comparison results are depicted in (a) to (g). The results prove that the proposed model can achieve high accuracy compared to the simulation experiments.

## 6. Analysis Based on the Proposed Model

In this section, the effect of path heterogeneity on performance of TCP flow transferred over multiple heterogeneous paths is analyzed based the proposed model. Firstly, the influence of Average Delay Asymmetry as well as Average Bandwidth Asymmetry on the throughput is investigated. Then we discuss the policy of determining appropriate number of links to transmit the segments of TCP flow over multiple heterogeneous paths.

### 6.1. The Influence of Delay and Bandwidth Asymmetry

It is an interesting issue to study to what extent do Average Delay Asymmetry and Average Bandwidth Asymmetry affect the throughput of TCP flow transferred over multiple heterogeneous paths. It has been previously concluded in section IV that the performance of TCP over multiple heterogeneous links is subject to these two parameters, but which is the main factor that affects the TCP performance, Average Delay Asymmetry or Average Bandwidth Asymmetry?

To answer this question, we use the proposed performance analysis model to evaluate the TCP throughput as a function of both Average Delay Asymmetry and Average Bandwidth Asymmetry. The minimum delay and bandwidth are 5ms and 100kbps. The Average Delay Asymmetry and Average Bandwidth Asymmetry are set to vary from 0ms to 35ms and from 0kbps to 700kbps. In this case, the maximum of Average Delay Asymmetry and Average Bandwidth Asymmetry are both seven times of minimum delay and bandwidth. The number of links utilized for concurrently transmitting data varies from 1 to 4. The evaluation results are shown in Figure 7.

Figure 7 shows that the average throughput drops significantly to the axis of Average Delay Asymmetry but decreases at a much slower pace to the axis of Average Bandwidth Asymmetry. This phenomenon is particularly obvious when four links are used to concurrently transfer the TCP flow. In this case, under highest level of Average Delay Asymmetry, the average throughput decreases by 1.8 times as the Average Bandwidth Asymmetry varies from zero to maximum. In contrast, when Average Bandwidth Asymmetry remains at highest level, and the Average Delay Asymmetry varies from zero to maximum, the average throughput is reduced by 2.8 times.

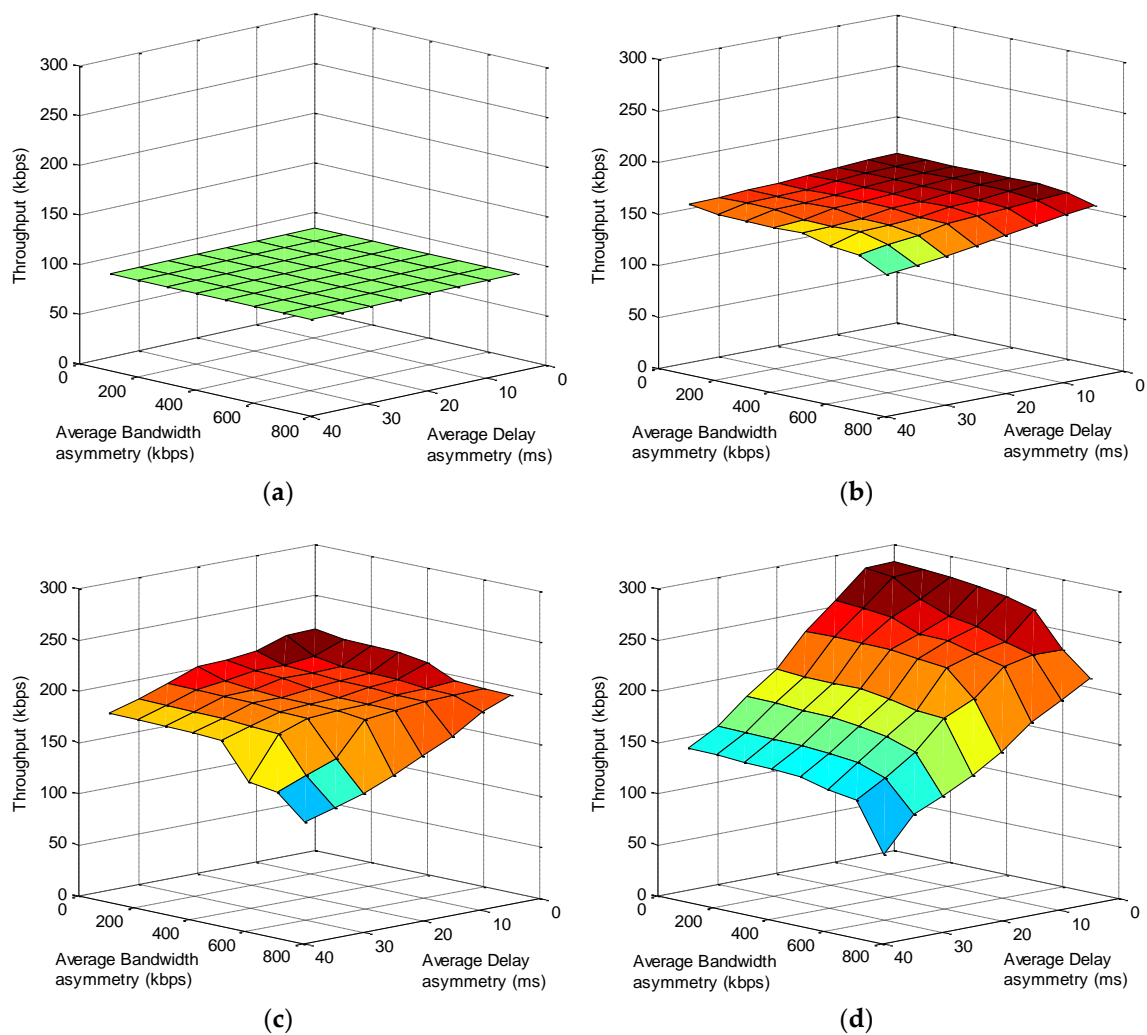

**Figure 7.** The throughput of TCP over multiple heterogeneous links on axes of both Average Bandwidth Asymmetry and Average Delay Asymmetry. The minimum delay is 5ms and the minimum bandwidth is 100kbps. (a), (b), (c) and (d) are the results using 1,2, 3 and 4 links. It is shown that the throughput is more prone to the effect of Average Delay Asymmetry.

Based on the above analysis, we can conduct that the Average Delay Asymmetry is the main factor that affects the throughput performance of TCP flow over multiple heterogeneous paths. This inference can guide the design of multipath transmission mechanism in heterogeneous networks.

### 6.2. Relationship Between the Throughput Performance and the Number of Links

Knowing that Average Delay Asymmetry is the dominant factor that affects the TCP throughput transferred over multiple heterogeneous paths, we can now investigate the relationship between the TCP performance and the number of links employed for transmission under different level of Average Delay Asymmetry. Further, the optimal number of links should be used to achieve optimized performance is discussed.

Under four groups of minimum delay, we evaluate the throughput of TCP flows employing different number of links as a function of Average Delay Asymmetry. During the evaluation, up to 4 links are utilized and the bandwidth of each link is 100kbps. Along with the Average Delay Asymmetry varying from 10ms to 90ms, the throughput is derived using the proposed model under the minimum delay of 5ms, 20ms, 35ms and 50ms respectively. The results of the evaluation are depicted in Figure 8.

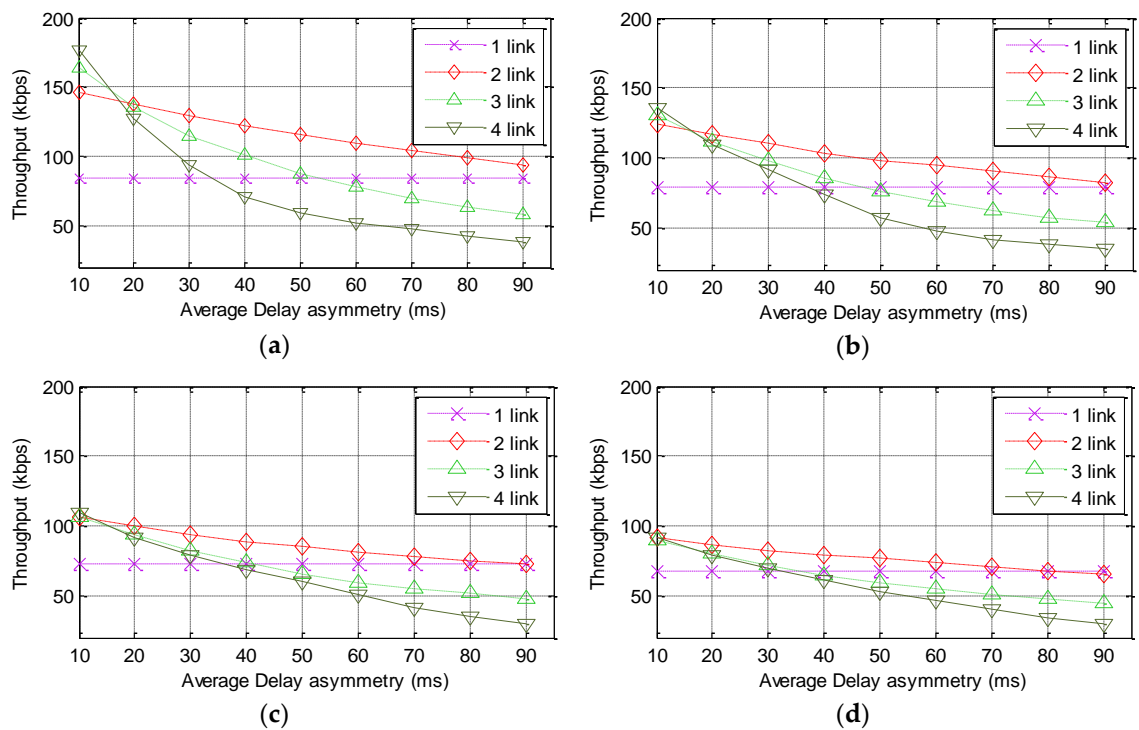

**Figure 8.** The throughput of TCP over multiple heterogeneous links as function of Average Delay Asymmetry using 1, 2, 3 and 4 links. (a), (b), (c) and (d) are the results with the minimum delay of 5ms, 20ms, 35ms and 50ms. It is shown that Average Delay Asymmetry compromises the benefits of aggregating bandwidth by utilizing multiple links.

According to Figure 8, we can find that the Average Delay Asymmetry of heterogeneous networks compromised the benefits of aggregating bandwidth by utilizing multiple links. Meanwhile, large minimum delay exacerbates the effect of Average Delay Asymmetry on throughput performance using multiple links. Under the minimum delay of 5ms, when the Average Delay Asymmetry increases to 35.6ms, the throughput of TCP concurrently transferred over four links decreases to that of TCP using only one link .When minimum delay increases from 5ms to 50ms, such threshold of Average Delay Asymmetry at which the throughput of four links equals to that of one link decreases from 35.6ms to 30.6ms.

Based on the above evaluation results, we can roughly derive a criterion of determining the number of links to optimize the throughput performance. For example, when the minimum delay is more than 5ms and the Average Delay Asymmetry is more than 20ms, utilizing two links to transfer the TCP flow will achieve maximum throughput.

## 7. Conclusion

In this paper, the severe extent of link quality asymmetry in real world situations is revealed based on field measurement, and then a performance analysis model for TCP over multiple heterogeneous paths for 5G services is derived regarding average throughput. Taking into the consideration of both bandwidth and delay asymmetry, we carefully investigate the transmission of TCP segments over multiple heterogeneous links and derive the corresponding performance analysis model. The proposed model is validated by comparison with simulation experiment using parameters from the field measurement. The results prove that the proposed performance analysis model can achieve high analytical accuracy in practical environment. Further analysis based on the proposed model reveals some interesting inferences. First, compared to bandwidth asymmetry, delay asymmetry is the dominant factor that affects the performance of TCP over heterogeneous networks. Second, the criteria of determining appropriate number of links to be used to optimize the TCP multipath performance is discussed. The proposed model can provide a guidance to the design of CMT solutions for 5G mobile services.

## 8. Acknowledgement

This work was supported by the Beijing Municipal Natural Science Foundation under Grant No. 4182048 and NSAF under Grant No. U1530118. An earlier version of this paper was presented at 2017 IEEE Global Communications Conference (GLOBECOM 2017).

## References

1.  Ramaboli, A. L.; Falowo, O. E.; Chan, A. H. Bandwidth aggregation in heterogeneous wireless networks: A survey of current approaches and issues. *J. Netw. Comput. Appl.* **2012**, 35, 1674–1690.
2.  Ford, A.; Raiciu, C.; Handley M.; Bonaventure, O. TCP extensions for multipath operation with multiple addresses. *IETF RFC 6824* 2013.
3.  Raiciu, C.; Paasch, C.; Barre, S.; Ford, A.; Honda, M.; Duchene, F.; Handley, M. How hard can it be? designing and implementing a deployable multipath TCP. In Proceedings of 9th USENIX conference on Networked Systems Design and Implementation, San Jose, CA, USA, 25-27 April 2012; pp. 29-29.
4.  Zhou, D.; Song, W.; Shi, M. Goodput improvement for multipath TCP by congestion window adaptation in multi-radio devices. In Proceedings of IEEE Consumer Communications and Networking Conference (CCNC), Las Vegas, NV, USA, 11-14 Januray 2013; pp. 508-514.
5.  Przybylski, M.; Belter, B.; Binczewski, A. Shall we worry about packet reordering. *Comput. Methods Sci. Technol.* **2005**, 11, 141–146.
6.  Bennett, J.; Partridge, C.; Shectman, N. Packet reordering is not pathological network behavior. *IEEE/ACM Trans. Netw.* **1999**, 7, 789–98.
7.  Chebrolu, K.; Rao, R. R. Bandwidth aggregation for real-time applications in heterogeneous wireless networks. *IEEE Trans. Mobile Comput.* **2006**, 5, 388-403.
8.  Kuhn, N.; Lochin, E.; Mifdaoui, A.; Sarwar, G.; Mehani, O.; Boreli, R. DAPS: Intelligent delay-aware packet scheduling for multipath transport. In Proceedings of IEEE International Conference on Communications (ICC), Sydney, NSW, Australia, 10-14 June 2014; pp. 1222-1227.
9.  Stewart, R. Stream control transmission protocol. *IETF RFC 4960* 2007.
10. Stewart, R.; Ramalho, M.; Xie, Q. Stream control transmission protocol (SCTP) partial reliability extension. *IETF RFC 3758* 2004.
11. Huang, C. M.; Tsai, C. H. WiMP-SCTP: Multi-path transmission using stream control transmission protocol (SCTP) in wireless networks. In Proceedings of Advanced Information Networking and Applications Workshops (AINAW), Niagara Falls, Ont., Canada, 21-23 May 2007; pp. 209–214.
12. Liao, J.; Wang, J.; Zhu, X. cmpSCTP: An extension of SCTP to support concurrent multi-path transfer. In Proceedings of IEEE International Conference on Communications (ICC), Beijing, China, 19-23 May 2008; pp. 5762-5766.
13. Ford, A.; Raiciu, C.; Handley, M.; Bonaventure, O. TCP extensions for multipath operation with multiple addresses. *IETF RFC 6824* 2013.
14. Anipko, D. Multiple Provisioning Domain Architecture. *IETF RFC 7556* 2015.
15. Zhang, H.; Zhang, Q.; Du, X. Toward Vehicle-Assisted Cloud Computing for Smartphones, IEEE Transactions on Vehicular Technology, Issue 12, Vol.64, pp. 5610-5618, Dec. 2015.
16. Du, X.; Zhang, M.; Nygard, K.; Guizani, S.; Chen, H. Self-Healing Sensor Networks with Distributed Decision Making, International Journal of Sensor Networks, Vol. 2, Nos. 5/6, pp. 289 –298, 2007.
17. Wu, L.; Du, X.; Wu, J. MobiFish: A Lightweight Anti-Phishing Scheme for Mobile Phones, in Proc. of the 23rd International Conference on Computer Communications and Networks (ICCCN), Shanghai, China, August 2014.

18. Xiao, Y.; Du, X.; Zhang, J.; Hu, F.; Guizani, S. Internet Protocol Television (IPTV): the Killer Application for the Next Generation Internet, IEEE Communications Magazine, Vol. 45, No. 11, pp. 126–134, Nov. 2007.

19. Zhang, H.; Chen, S.; Li, X.; Ji, H.; Du, X. Interference Management for Heterogeneous Network with Spectral Efficiency Improvement, IEEE Wireless Communications Magazine, Issue 2, Vol. 22, pp. 101-107, April 2015.

20. Du, X.; Xiao, Y.; Guizani, M.; Chen, H. An Effective Key Management Scheme for Heterogeneous Sensor Networks, Ad Hoc Networks, Elsevier, Vol. 5, Issue 1, pp 24–34, Jan. 2007.

21. Lee, C.; Chuang, M. Seamless handover for high-speed trains using femtocell-based multiple egress network Interfaces. *IEEE Trans. Wireless. Commun.* **2014**, 13, 6619-6628.

22. Huang, L.; Zhu, G.; Du, X. Cognitive Femtocell Networks: An Opportunistic Spectrum Access for Future Indoor Wireless Coverage. *IEEE Wireless Commun. Mag.* **2013**, 20, 44-51.

23. Zhao, J.; Xu, C.; Guan, J.; Zhang, H. A fluid model of multipath TCP algorithm: Fairness design with congestion balancing. In Proceedings of IEEE International Conference on Communications (ICC), London, UK, 8-12 June 2015; pp. 6965-6970.

24. Du, X.; Lin, F. Maintaining Differentiated Coverage in Heterogeneous Sensor Networks. *EURASIP J. Wireless Commun. Netw.* **2005**, 5, 565-572.

25. Zhang, M.; Du, X.; Nygard, K. Improving Coverage Performance in Sensor Networks by Using Mobile Sensors. In Proceedings of IEEE Military Communications Conference (MILCOM), Atlantic City, NJ, USA, 17-20 October 2005; pp. 1-6.

26. Jayasumana, A.; Piratla, N.; Banka, T.; Bare, A.; Whitner, R. Improved packet reordering metrics. *IETF RFC 5236* 2008.

27. Kaspar, D. Multipath aggregation of heterogeneous access networks. *ACM SIGMultimedia Records* **2012**, 4, 27-28.

28. Paasch, C.; Ferlin, S.; Alay, O.; Bonaventure, O. Experimental evaluation of multipath TCP schedulers. In Proceedings of the 2014 ACM SIGCOMM workshop on Capacity sharing, Chicago, IL, USA, 18-18 August 2014; pp.27 -32.

29. Partov, B.; Leith, D. J. Experimental evaluation of multi-path schedulers for LTE/wifi devices. In Proceedings of ACM International Workshop on Wireless Network Testbeds, Experimental Evaluation, and Characterization (WiNTECH), New York City, NY, USA, 3-7 October 2016; pp. 41-48.

30. Benet, C. H.; Kassler, A.; Zola, E. Predicting Expected TCP throughput using genetic algorithm. *Comput. Netw.* **2016**, 108, 307-322.

31. Parvez, N.; Mahanti, A.; Williamson, C. An Analytic Throughput Model for TCP NewReno. *IEEE/ACM Trans. Netw.* **2010**, 18, 448-461.

32. Romero-Angeles, R. I.; Garcia-Ruiz, R.; Lara-Rodriguez, D. An Algorithm for the Evaluation of the Throughput of a TCP NewReno Bulk Data Flow. *IEEE Commun. Lett.* **2015**, 19, 941-944.

33. Alrshah, M. A.; Othman, M.; Ali, B.; Hanapi, Z. Modeling the Throughput of the Linux-based Agile-SD Transmission Control Protocol. *IEEE Access* **2017**.

34. Panda, M.; Vu, H. L.; Mandjes, M.; Pokhrel, S. R. Performance Analysis of TCP NewReno over a Cellular Last-Mile: Buffer and Channel Loss. *IEEE Trans. Mobile Comput.* **2015**, 14, 1629-1643.

35. Li, M.; Lukyanenko, A.; Ou, Z. Multipath transmission for the internet: A survey. *IEEE Commun. Surveys Tuts.* **2016**, 18, 2887-2925.

36. Henderson, T.; Floyd, S.; Gurtov, A.; Nishida, Y. The NewReno Modification to TCP's Fast Recovery Algorithm, *IETF RFC 6582* 2012.

37. The ns-3 Network Simulator. Available online: http://www.nsnam.org (accessed on 20 Mar. 2018).